\documentclass[amsmath,amssymb,floatfix,prb,epsf,showpacs,twocolumn]{revtex4}
\usepackage{amsmath,amssymb,natbib,bm,graphicx,url,epsfig}
\usepackage[ansinew]{inputenc}
\usepackage{bm}

\newcommand{\be}{\begin{equation}}
\newcommand{\ee}{\end{equation}}
\newcommand{\bes}{\begin{equation}\begin{split}}
\newcommand{\ees}{\end{split}\end{equation}}
\newcommand{\ba}{\begin{eqnarray}}
\newcommand{\ea}{\end{eqnarray}}
\newcommand{\nn}{\nonumber}

\newcommand{\RR}{\mathbb{R}}

 \DeclareMathOperator{\tr}{tr}

\def\beq{\begin{equation}}
\def\eeq{\end{equation}}
\def\bea{\begin{eqnarray}}
\def\eea{\end{eqnarray}}

\begin{document}

\title{ Boundary  Wess-Zumino-Novikov-Witten Model from the Pairing Hamiltonian
}
\author{Tigran A.~Sedrakyan and Victor Galitski}

\affiliation{Joint Quantum Institute and Condensed Matter Theory Center, Department of Physics, University of Maryland, College Park, MD 20742, USA}

\date{\today}

\begin{abstract}
Correlation functions in the Wess-Zumino-Novikov-Witten (WZNW) theory satisfy a system of Knizhnik-Zamolodchikov (KZ) equations,
which involve constants of motion of an exactly solvable model, known as Gaudin magnet.
We show that modified KZ equations, where the Gaudin operators are replaced by constants of motion of the
closely related pairing Hamiltonian, give rise to a deformed WZNW model that contains terms  breaking translational symmetry.
This boundary WZNW model is identified and solved. The solution establishes a connection between  the WZNW model and the
pairing Hamiltonian in the theory of superconductivity. We also argue and demonstrate on an explicit example that our general
approach can be used to derive exact solutions to a variety of dynamical systems.
\end{abstract}
\pacs{71.10.Pm, 74.20.Fg, 02.30.Ik}
\maketitle

\section{Introduction}

The Wess-Zumino-Novikov-Witten (WZNW) model plays an important role in 
physics. Historically, the $SU(2)$ version of the model with topological coupling $k=1$
was used to describe low-energy dynamics of a one-dimensional spin-$1/2$
Heisenberg antiferromagnet. At higher integer couplings, $k$, it describes
quantum critical points in the parameter space of quantum
antiferromagnetic {\em spin-$S$} chains, with $S=k/2$~\cite{Afl}.
Many other applications of the WZNW model have emerged in
various contexts lately (see, {\em e.g.}, Ref.~[\onlinecite{ludwig}]). The Lagrangian formulation of the theory
is given by a non-linear sigma-model defined in the Euclidean space
by the  action~\cite{PW}
\bea
\label{WZW}
S_{WZNW}(g)=&&\!\!\!\!\!\!\frac{k}{16\pi}\int\limits_{S^2} dz d\bar{z} \tr \left[ \partial_{a}g^{\dagger}\partial^{a}g \right]\qquad\qquad \\
&-&\frac{i k}{24 \pi}\int\limits_{B^3} d^3 x \varepsilon ^{\mu\nu\rho}\tr\left[ g^{\dagger}\partial _{\mu}g
g^{\dagger}\partial _{\nu}g g^{\dagger}\partial _{\rho}g \right],\quad\nn
\eea
where integration in the second topological Wess-Zumino term is over a three-dimensional ball,
$x = (z,\bar{z},\xi) \in B^3$, whose boundary at $\xi = 0$ is the two-dimensional sphere, $S^2 = \partial B^3$,
which corresponds to a compactified complex plane parametrized by $(z,\bar{z})$ and $g(z,\bar{z}) \in SU(2)$.
The integer parameter $k$ in Eq.~(\ref{WZW}) is the level of the corresponding conformal field theory (CFT).
The WZNW action is invariant under conformal and non-Abelian current algebras. The current algebra
transformations have a chiral structure, {\em i.e.} they act on the group element $g(z,\bar{z})$ as
$g^{\prime}(z,\bar{z})= {\cal{U}}(z)g(z,\bar{z})\bar{\cal{U}}(\bar{z})$. Here ${\cal{U}}(z)$ and $\bar{\cal{U}}(\bar{z})$ are independent
elements of the group $SU(2)$. This property allows to study the holomorphic ($z$-dependent) and antiholomorphic ($\bar{z}$-dependent)
sectors of the model separately (below, we focus on the holomorphic sector).

The $N$-point correlators of primary fields, $G(z_1\ldots z_N) = \left\langle \phi(z_1,\bar{z}_1) \ldots \phi(z_N,\bar{z}_N) \right\rangle_{S_{\rm WZNW}}$,
satisfy the Knizhnik-Zamolodchikov (KZ) equations~\cite{KZ},
 \bea
\label{GKZ}
\left[ (k+2)\partial_{z_l} - \hat{ H}_l^G \right] G\left(\{z_l\}\right) = 0,
\eea
with
\bea
\label{GKZ-1}
\hat{ H}_l^G =
\sum\limits_{l \ne l'} w(z_l,z_{l'}) { \hat{{\bf S}}_l \cdot
\hat{{\bf S}}_{l'} }
\eea
$ l,l'=1,2,\ldots N$, and $w_{l,l'} = \left(z_l - z_{l'} \right)^{-1}$.
Here $\hat{{\bf S}}_{l}=(\hat{S}^1_l,\hat{S}^2_l,\hat{S}^3_l)$ and
$\hat{S}^a_l$ are SU(2) generators.
 Amazingly, operators $\hat{ H}_l^{G}$ in Eq.~(\ref{GKZ}) are formally equivalent to the integrals of motion of a seemingly unrelated
Gaudin magnet model~\cite{Gaudin,Sierra_RMP}. The Gaudin magnet represents a quantum spin Hamiltonian, with effective long-range interactions
between  spins, which is exactly solvable ``by design.'' Its Hamiltonian can be represented as a linear combination of the mutually commuting
integrals of motion, $\left[  \hat{ H}_l^{G}, \hat{ H}_{l'}^{G} \right] = 0, \forall l, l'$ as
follows:
\bea
\label{IM}
\hat{\cal H} = 2 \sum_l z_l\hat{ H}_l^{G}.
\eea

One general question that we formulate in this paper is whether it is possible to derive  deformed WZNW models, whose correlators satisfy
modified KZ equations (\ref{GKZ}) with  a different set of  operators $\hat{ H}_l$. Below, we answer this question in the affirmative by
providing an example of this ``reverse engineering approach'' and finding a boundary WZNW model, which corresponds to the operators $\hat{ H}_l$
representing the integrals of motion of the discrete pairing Hamiltonian
(Richardson model)~\cite{R1, Delft1, italians, Amico, Delft2, Sierra1, Sierra_RMP,teodorescu, Yuzbash,caux,amico2,garcia} closely related to
the Gaudin magnet. It descends from the familiar BCS Hamiltonian
\bea
\label{R}
\hat{\cal H}_{\rm RBCS}=\sum_{l,s=\pm} z_l \hat{c}^\dagger_{l s} \hat{c}_{l s}- \lambda
\sum_{l,l^{\prime} S}\hat{c}^\dagger_{l +} \hat{c}^\dagger_{l -} \hat{c}_{l' -} \hat{c}_{l' +},
\eea
where $\hat{c}^\dagger_{l s}$ and $\hat{c}_{l s}$ are fermion creation/annihilation  operators corresponding to a
single-particle state $\left| l \right\rangle$ with energies $z_l$ and spin $s = \pm$. If  $\lambda > 0$, the ground state is a superconductor with
all fermions paired~\cite{Ralph}.  Then, the operators $\hat{c}^\dagger_{l +} \hat{c}^\dagger_{l -}$, $\hat{c}_{l' -} \hat{c}_{l' +}$, and
$\sum_s \left( \hat{c}^\dagger_{l s} \hat{c}_{l s} - 1/2\right)/2$ become algebraically equivalent
to the Pauli matrices $\hat{\sigma}^\dagger_l$, $\hat{\sigma}^-_l$, and $\hat{\sigma}^3_l$ (Anderson pseudospins). The corresponding spin
Hamiltonian is the integrable Richardson model, which can be presented in two identical ways
\begin{eqnarray}
\label{R1}
\hat{\cal H}_{\rm Rich}&=&\sum_{l}^N z_l(1 + \hat{\sigma}_l^3)- {\lambda \over 4} \sum_{l,l'}\hat{\sigma}_l^\dagger \hat{\sigma}_{l'}^- \\
&\equiv& - \sum_l \left( 2 z_l \hat{H}_l^R - z_l + {\lambda \over 4} \right)
+\lambda \left(\sum_l \hat{H}_l^R\right)^2,\nn
\label{R2}
\end{eqnarray}
where $\hat{\sigma}^{3,\pm}$ are Pauli matrices, and the operators $\hat{H}_l^R = -\hat{\sigma}^3_l/2 + \lambda \hat{H}_l^G$ represent $N$
mutually commuting $\left[ \hat{H}_l^R, \hat{H}_{l'}^R\right] = 0$ conserved ``currents.'' Note that  $\sum_l \hat{H}_l^G =0$, and hence the second
term in Eq.~(\ref{R2}) can be simplified as
$\sum_l \hat{H}_l^R = - \sum_l \hat{\sigma}^3_l/2$ to give the ``total pseudo-spin magnetization,'' which separates  the Hilbert space into sectors
with different numbers of Cooper pairs, which were actually studied in Ref.~\onlinecite{R1}.

Note that the  Gaudin model is closely related to the Richardson model (\ref{R1},\ref{R2}) and corresponds to its infinite coupling limit.
It is interesting to see what perturbed WZNW model, would correspond to the KZ equations~(\ref{GKZ}) with the operators $\hat{H}_l$ replaced with
$\hat{H}_l^R$. This is a key question addressed in this paper, but in the interest of practical applications, we shall consider a more general form
of ``new'' operators:
\bea
\label{rot}
&&\hat{\tilde{H}}_l \left[{\cal C} \right] = - \hat{U}[{\cal C}]\partial _{z_l}\hat{U}^{-1}[{\cal C}]
+\lambda \hat{\tilde H}_l^G[{\cal C}],\;\;\;\;\;{\text{where}}\nn\\
&&\hat{\tilde H}_l^G[{\cal C}]= {1 \over k+2}\,\,
\hat{U}[{\cal C}] \,
\hat{H}_l^G\, \hat{U}^{-1}[{\cal C}] .
\eea
Here
 $\hat{\tilde H}_l^G[{\cal C}]$ is a rotated Gaudin Hamiltonian, with $\hat{U}[{\cal C}] = \exp\left\{\sum_i q(z_i){\hat S}^3_i
\Theta\left[{\cal C}, z_i\right]\right\}$, ${\cal C}$ is a closed contour in the complex plane, $q(z)$ is an arbitrary analytic function
inside $\mathcal{C}$, and $\Theta\left[{\cal C}, z\right] = 1$ if $z$ lies within the region
enclosed by the contour and zero otherwise.
We emphasize that Eq.~(\ref{rot}) contains the  conserved ``currents'' from the Richardson pairing model (\ref{R1}) in the simplest special case,
of $k=1$ (i.e., $\hat{\bf S}$ become Pauli matrices $\hat{\bm \sigma}/2$), ${\cal C}\rightarrow {\cal C}_\infty$ (i.e., the contour ${\cal C}_\infty$ encloses
all points in $\mathbb{C}$), and $q(z) = -z/\lambda$, so that  $\hat{U}_{\rm Rich} = \exp\left\{-\sum_i z_i{\hat \sigma}^3_i/(2 \lambda) \right\}$.

We first present the main result for the boundary WZNW model, corresponding to operators $\hat{\tilde{H}}_l \left[{\cal C} \right]$ defined in Eq.~(\ref{rot}),
\bea
\label{action}
S_{\rm BWZNW} \left[ {\cal C} \right]= S_{\rm WZNW} +S_{bound}^L[{\cal C}]+S_{bound}^R[\overline{\cal C}],
\eea
where $ S_{\rm WZNW}$ is the standard WZNW action~(\ref{WZW}),
\bea
\label{bound}
S_{bound}^L[{\cal C}]=-\oint_{C} dz q(z)  J^3(z)
\eea
is the ``left'' boundary term, and the ``right'' boundary term, $S_{bound}^R[\overline{\cal C}]$ is given by (\ref{bound}) with
$z\rightarrow \bar{z}$ and $J^3(z) \to \bar{J}^3(\bar{z})$.
In Eq.~(\ref{bound}), $J^3(z)$ is a component of the ``left'' current in the $SU(2)$ WZNW theory, defined in a standard way:
$J^a(z)=(k/2)\tr \left[\hat{S}^a g(z,\bar{z})\partial_zg^{\dagger}(z,\bar{z})\right]$, $a=\pm, 3$. Note that due to conformal invariance the ``left''
currents do not depend on $\bar{z}$ and likewise the ``right'' currents, $\bar{J}^a(\bar{z})$ do not depend on $z$. Note that the term (\ref{bound})
breaks translational invariance of the model and hence can be interpreted as a generalized impurity~\cite{ludwig}.
Below we  prove that the boundary action gives rises to generalized KZ Eqs.~(\ref{GKZ}) and present exact results for the corresponding correlation functions.

\section{Derivation of the boundary WZNW action}

We are seeking to prove that a correlation function of arbitrary primary fields~\cite{CFT} in
the $SU(2)$ boundary WZNW model (\ref{action}),
\bea
\label{GG}
G(z_1,\cdots z_N)&=&\langle \phi_{s_1}(z_1)\ldots \phi_{s_N}(z_N)\rangle _{S_{\rm BWZNW}} \nn\\
&\equiv&
\langle \Phi\left[{\cal{C}}\right] \phi_{s_1}(z_1)\cdots \phi_{s_N}(z_N) \rangle _{S_{\rm WZNW}}
\eea
satisfies the generalized KZ equations (\ref{GKZ})
with operators (\ref{rot}). Here, $\Phi\left[{\cal C}\right]=e^{-S_{bound}({\cal C})}$  and $s_i$ stands for the spin, $0\leq s_i\leq (k/2)$,
$i=1\ldots N$.


To solve the KZ equations (\ref{GKZ}) we look for $\Phi\left[{\cal C}\right]$ in the form (\ref{bound})
$ \Phi\left[{\cal C}\right]= e^{\oint_{C} dz q(z)  J^3(z)}$, where $q(z)$ is an analytic and differentiable function in $\mathcal{C}$,
and utilize the two standard key ingredients of the $SU(2)$ WZNW theory and CFTs~\cite{CFT}:
{\em (i)} The crux here is the operator product expansion satisfied by the currents with the same chiralities
(currents with different chiralities commute); {\em (ii)}  Action of the Virasoro generators on primary fields.
Then the expression for the correlation function $G(z_1,\cdots z_N)$ can be simplified by contracting $J^z(z)$ in
$\Phi[{\cal C}]=\sum_p\bigl(1/p!\bigr) \Big(\oint_{\cal C} dz q(z)  J^3(z)\Big)^p$ with all primary fields
\begin{eqnarray}
\label{ope21}
G(z_1,\cdots z_N)=\Bigl\langle e^{\sum_i q(z_i)  \hat S^3_i\Theta \left[{\cal C}, z_i\right]}
\phi_{s_1}(z_1)\cdots \phi_{s_N}(z_N) \Bigr\rangle,\nn\;\;\;
\end{eqnarray}
where
the functional averaging with respect to the ${S_{\rm WZNW}}$ is understood.
Using  the standard technique of Ref.~\onlinecite{KZ} and taking into  account the boundary operator we find that indeed the following
identity holds
\bea
\label{KZ2}
 \Bigg[\partial_{z_i}-q^{\prime}(z_i)\hat {S}^3_i\Theta \left[{\cal C}, z_i\right]
- \hat{\tilde H}_i^G[{\cal C}]\Bigg ]G(z_1,\cdots z_N)=0, \qquad
\eea
where $\hat{\tilde H}_i^G[{\cal C}]$
is the rotated Gaudin Hamiltonian defined in (\ref{rot}). Note that if $q(z)\equiv - z/k \lambda$ and all $z_i$ $(i=1 \ldots N)$
are inside ${\cal{C}}$, Eq.~(\ref{KZ2}) precisely reproduces modified KZ equations (\ref{GKZ}) with the Gaudin integrals of motion replaced with
those of the Richardson model with the interaction parameter $\lambda$. Hence, we recover the amazing fact that the correlation functions
of the boundary WZNW model carry information about the exact correlation functions of the pairing model.

\maketitle
\section{Solution of the Knizhnik-Zamolodchikov equations}
We now show that the generalized KZ Eqs.~(\ref{GKZ}) can be solved exactly using the standard off-shell Bethe Ansatz
technique~\cite{Babujian}. This method applies if all $\hat{\tilde H}_l$ are commuting, in what follows we will consider
the case when all $z_i\in {\cal C}$.
Look for the solution
in an integral form
\bea
\label{solution}
G(z_1 \cdots z_N)=\oint \prod_{k=1}^M du_k
\chi(\{u_{\alpha}\}|\{z_i\})
{\cal V}(\{u_{\alpha}\}|\{z_i\}),\;\;\;
\eea
where $M$ is fixed from the condition that the correlation function $G(z_1 \cdots z_N)$ should be a
singlet with respect to the global $SU(2)$:
$\hat S^3G(z_1 \cdots z_N)\equiv 0$ for WZNW model~\cite{KZ} and also for our boundary case.
The integrations are to be taken here over canonical cycles in the n dimensional complex space
where points $z_i$ are excluded, with coefficients, defined by
the monodromy group of the function $\chi(\{u_{\alpha}\}, \{z_i\})$.
Then it follows that $M=\sum_{i=1}^N s_i$.
We look for eigenstates of a set of commuting Hamiltonian operators, $\hat{H}_i^R$, of the pairing model
in the form
\bea
\label{sol-pm}
{\cal V}(u_1\cdots u_M |\{z_i\})=\hat S^\dagger(u_1)\cdots \hat S^\dagger(u_M)\vert 0 \rangle,
\eea
where
\bea
\label{VS}
\hat S^\dagger(u)=\sum_{i=1}^{N}\frac{\hat S_i^\dagger}{u-z_i}
\eea
 and the bare vacuum state, $\vert 0\rangle$,
is a direct
product of lowest weight vectors of the corresponding representation $s_i$:
$\hat S_i^3\vert s_i,m_i \rangle=m_i\vert s_i,m_i \rangle $, where $m_i=-s_i$. For example if $k=1$ and all
primary fields in $G(z_1 \cdots z_N)$ are from $s_i=1/2$ representation space of
SU(2), then $M=N/2$ and $\vert 0 \rangle =\left(\begin{array}{c}
                               0 \\
                               1
                             \end{array}\right)_1\otimes\ldots\otimes\left(\begin{array}{c}
                               0 \\
                               1
                             \end{array}\right)_N.
$
In the basis where the primary fields are defined by spin $s$ and its
z-projection, $m=-s \ldots s$, their correlation function
$\langle \phi^{m_1}_{s_1}(z_1) \cdots \phi^{m_N}_{s_N}(z_N) \rangle$
is connected to the general expression (\ref{solution})
as follows
\bea
\label{solution2}
&&\langle \phi^{m_1}_{s_1}(z_1) \cdots \phi^{m_N}_{s_N}(z_N) \rangle \nn\\
 &=& \langle s_N, m_N|\cdots \langle s_1, m_1|G(z_1\cdots z_N),\qquad
\eea
where $\langle s_i, m_i \mid \hat S^3_i= \langle s_i, m_i \mid m_i $.
We found that 
$\chi(\{u_{\alpha}\}|\{z_i\})$ has the following form
\bea
\label{sol3}
&&\chi(\{u_{\alpha}\}|\{z_i\})
=\chi_0(\{u_{\alpha}\}|\{z_i\})\qquad \\
&\times& \exp\Bigg\{\frac{1}{\lambda (k+2)}\left[-\lambda k \!\sum_{i=1}^N m_i q(z_i)\!+\!\!\sum_{\alpha} u_{\alpha}\right]\Bigg\},\nn
\eea
where
\bea
\label{sol0}
\nonumber
\chi_0(\{u_{\alpha}\}|\{z_i\})&=&\prod_{i\neq j}(z_i-z_j)^{m_i m_j \over(k+2)}
\prod_{\beta \neq \alpha}(u_{\alpha}-u_{\beta})^{1 \over (k+2)}\nn\\
&\times& \prod_{\gamma, i}(u_{\gamma}-z_i)^{-m_i \over (k+2)}
\eea
is the known solution~\cite{Babujian,gogolin} to the KZ Eqs. for the canonic $SU(2)$ WZNW model.
In general these solutions can be expressed  analytically
in terms of multi-variable confluent hypergeometric functions \cite{Aomoto, Gelfand}.
Note that
when $\lambda\rightarrow\infty$, the Richardson pseudospin model reduces to the Gaudin magnet,
and consequently $\chi$ in Eq.~(\ref{sol3}) reduces to $\chi_0$.
  Moreover, analysis of Eqs.~(\ref{solution}) at $q(z)\equiv -z/k\lambda$ suggests that the integral over
$u_1\ldots u_M$ in Eq.~(\ref{solution}) has a
saddle point  defined by the condition
\bea
\label{BANZ}
\frac{1}{\lambda}+\sum_{\alpha \neq \beta}^M \frac{1}{u_{\alpha}-u_{\beta}}
+\sum_{i=1}^N  \frac{s_i}{u_{\beta}-z_i}=0.
\eea
Interestingly,
this condition coincides with Richardson equations for the eigenvalues of the reduced BCS Hamiltonian
(\ref{R1}).


\section{Bosonized action at $k=1$: Implications}

The WZNW model (\ref{action}) at $k=1$  can also be realized as  a free boson theory
with central charge $c=1$~\cite{gogolin}. Following  the standard bosonization technique we introduce a scalar field,
$\hat\varphi=\hat\varphi(z)+\hat{\bar\varphi}(\bar{z})$,
and rewrite the z-component of the current in the form
$
J^3(z)=\frac{i}{\sqrt{2 \pi}} \partial_z \hat\varphi(z).
$
Note that this representation of $J^z$ is correct only locally. The full action (\ref{WZW}) at $k=1$ reads
\bea
\label{action2}
S=\frac{1}{4\pi} \int d z d \bar{z} \partial_z \varphi \partial_{\bar z} \varphi + \frac{i}{ \lambda \sqrt{2 \pi}}
\left[\oint_{\cal C} d z z \partial_z\varphi + {\rm a.~c.}\right],\;\;\;
\eea
where ${\rm a.~c.}$ stands for an antiholomorphic contribution. We note that this bosonized version of the $k=1$ action was discussed  earlier
in Refs.~\onlinecite{Sierra1, Sierra2} by Sierra.
\begin{figure}[t]
\centerline{\includegraphics[width=85mm,angle=0,clip]{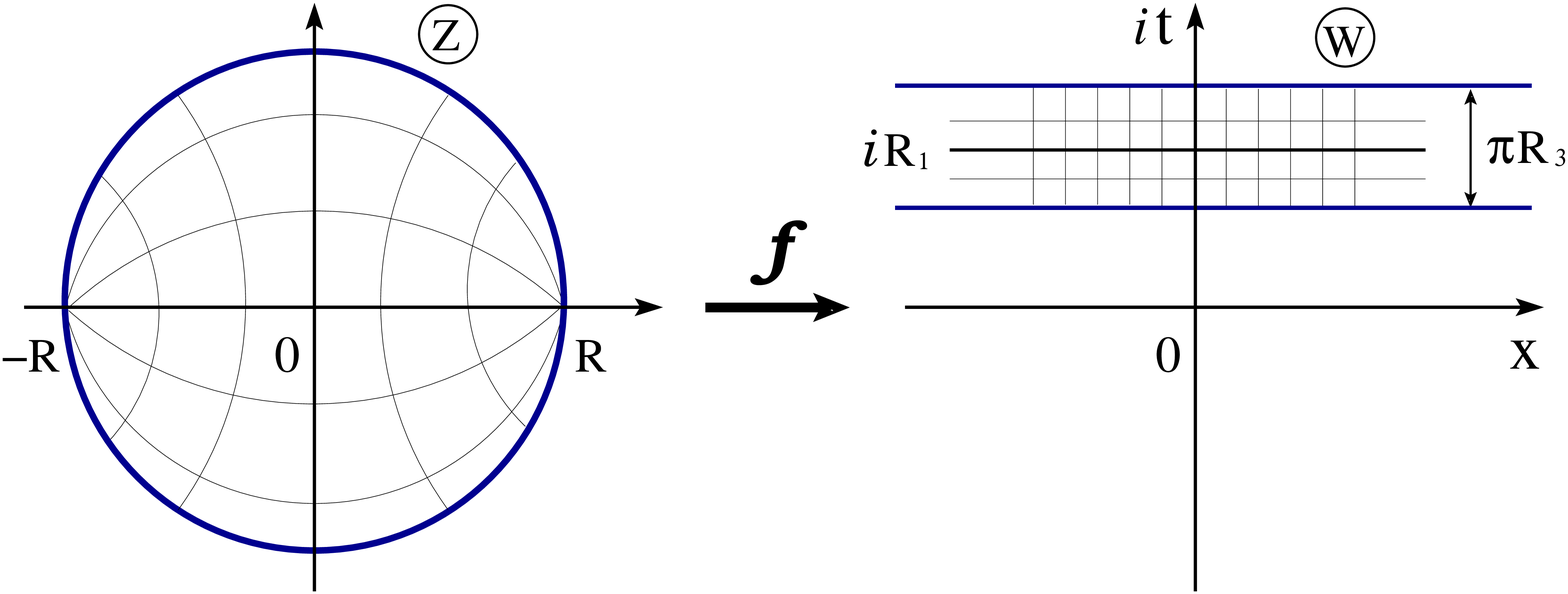}}
\caption{(Color online) An example of a conformal map.
Function $w=f(z)=iR_1+R_2+R_3\log\left(\frac{z-R}{z+R}\right)$ maps a disc with radius $R$ to a strip of width  $\pi R_3$ centered at $Im(w)=R_1$.}
 \label{Cond}
\end{figure}

Here we emphasize that due to the presence of the boundary term defined by an arbitrary contour ${\cal C}$
and  conformal invariance of the WZNW model make the boundary WZNW model a very useful tool to classify
and study low-energy, strong coupling disordered quantum systems~\cite{Konik} as well as various systems driven out of equilibrium.
Let us use bosonic version of the $k=1$ theory as an illustrative example.
If all $z_i$ are real, we have a standard physical Richardson pseudo-spin model. Here contour ${\cal C}$ can
be chosen as boundary of a narrow strip encompassing all $z_i$, which explicitly shows that we have an  equilibrium
system. On contrary, if some of $z_i$ have nonzero imaginary part, the contour
can be a circle with radius $R$. In this case parametrization $z=it+x$, where
$t$ is the dimensionless time and $x$ is the dimensionless coordinate, is
inconvenient, as we generate a complicated time dependent term in our bosonic
Hamiltonian. Interestingly enough, this, from a first sight abstract problem is
closely related to another, well defined and physically motivated system.
Consider the conformal map, $z\rightarrow w$, where $w=it+x$,
which transforms a disc with radius $R$ to an infinite strip,
see Fig.~1. The bulk action is invariant under such transformations, while
the boundary term will transform into $\sim\oint _{{\cal C}^\prime}dw\sinh^{-2}\left(\frac{w-iR_1-R_2}{2R_3}\right)\varphi\left[z(w)\right]$.
 Here contour ${\cal C}^\prime$ is the boundary of a strip of width  $\pi R_3$ centered at $t_0=R_1$. This term
 contributes to the Hamiltonian and makes it time dependent. It describes a single instantaneous perturbation
on the system at $t=t_0$, which however does not brake integrability.
Remarkably, we can extract enormous information about physical properties of such
systems by  analyzing exact correlation functions.


\maketitle
\section{Practical applications of the boundary WZNW model}

Boundary action Eq.~(\ref{bound}) together with the expression (\ref{sol3}) for the correlation
functions represent our main mathematical result. As argued, it has important consequences for a variety of seemingly
unrelated physical models, notably dynamical systems. We provide here an explicit example
of such correspondence between  Maxwell-Bloch (MB) theory of a two-level laser~\cite{lasers}, which is shown
to map onto the BWZNW model.  Below, we derive for the first time an exact solution to the system of MB equations with damping
\begin{eqnarray}
\label{MBEqs}
\nonumber
&&\partial_{\eta}{\mathcal E}+\gamma {\mathcal E}={\mathcal P},\\
&&\partial_{\xi} {\mathcal P}+\gamma_{\perp} {\mathcal P}={\mathcal N}{\mathcal E},\\
&&\partial_{\xi}{\mathcal N}+ {1 \over 2} ({\mathcal E}{\mathcal P}^*+{\mathcal E}^* {\mathcal P})=-\gamma_{||}{\mathcal N} + {\mathcal N}_0,
\nonumber
\end{eqnarray}
where ${\mathcal E}(\eta,\xi)$ is the complex electrical field amplitude, ${\mathcal P}(\eta,\xi)$ is the polarization of the medium,
${\mathcal N}(\eta,\xi)$ is the population inversion, and $\eta=\Omega x/c$ and $\xi=\Omega(t-x/c)$ are given in terms of real
space, $x$, and time, $t$, with $\Omega$ being a physical constant that depends on material
and cavity medium, and $c$ is the speed of light. In Eq.(\ref{MBEqs}), $\gamma \geq 0$ is a decay rate of energy losses inside the
laser medium and the constants $\gamma_\perp$ and $\gamma_{||}$ are damping coefficients of medium polarization and population
inversion. Dissipation in the population inversion equation
tends to return ${\mathcal N}$ to ${\mathcal N}_0/\gamma_{||}$, which is determined by the pumping.

Amazingly, the Hamiltonian formulation of (\ref{MBEqs}) for $\gamma_{\|}=\gamma_{\bot}=\gamma=0$  reduces to a
set of KZ equations (\ref{KZ2}) with linear $q(z)\equiv \alpha z$~\cite{BZM, Kitaev}. We show now
that the BWZNW model with $q(z)\equiv \alpha z +\beta z^2$ and the corresponding generalized KZ Eqs.~(\ref{KZ2}) describe the
system (\ref{MBEqs}) with finite damping parameters $\gamma,  \gamma_{\perp} \geq 0$.
First, we observe by analogy with Ref.~[\onlinecite{BZM}] that the set of MB equations~(\ref{MBEqs})  with {\em damping and pumping}
can be obtained from the compatibility condition of the following system of liner differential
equations with complex spectral parameter $z\in \mathbb{C}$
\bea
\label{MB}
\partial_{\xi}\psi=\left[\left(z-\frac{\gamma_{\bot}}{2}\right)\sigma_3+U_0\right]\psi,
\eea
and
\bea
\label{MB1}
\Bigg(\partial_{\eta}+\frac{{\mathcal N}_0}{z}\partial_z\Bigg)\psi=
\Bigg(\frac{\hat \rho}{4 z}-\frac{\gamma}{2}\sigma_3\Bigg)\psi,
\eea
where
\bea
\label{MB22}
U_0=\frac{1}{2}\left(\begin{array}{cc}
0& {\mathcal E} \\
-{\mathcal E}^{*}& 0
\end{array} \right), \;\; \hat\rho=\frac{1}{2}\left(\begin{array}{cc}
{\mathcal N}& -{\mathcal P} \\
-{\mathcal P}^{*}& -{\mathcal N}
\end{array} \right).
\eea
According to the method of isomonodromy solutions of differential equations,~\cite{Its,Kitaev}
variety of solutions of the MB equations can be obtained
by classifying solutions of an auxiliary equation,
$\partial_z \psi(z)=A(z,\xi,\eta)\psi(z)$, that are
consistent with the original MB equations.
We found that consistent with (\ref{MB})
choice of $A(z,\xi,\eta)$, which produces $N$-soliton solutions of MB equations in the presence of
pumping and damping, reads
\bea
\label{A}
A(z,\xi,\eta)&=&(\xi-\xi_0)\sigma_3+ \sum_{j=1}^N \frac{A_j}{z-z_j},\nn\\
 z_j&=&\sqrt{2 N_0 \eta-k_j^2},
\eea
with parameters $k_j^2, \xi_0 \in \RR$. Substituting expression for $\partial_z \psi(z)$ together with
Eq.~(\ref{A})
into Eqs.~(\ref{MB}), writing compatibility conditions and
equating the residues of the poles at $z=z_i$, $i=1\ldots N$, one will
obtain for functions $A_j$:
\bea
\label{MB2}
\partial_{\xi}A_j=\Big[\Big(z_j-\frac{\gamma_{\bot}}{2}\Big)\sigma_3+U_0, A_j\Big],\nn\\
\partial_{\eta}A_j=\Big[\Big(\frac{\hat\rho}{4 z_j}-\frac{\gamma}{2}\Big)\sigma_3+U_0, A_j\Big]
\eea
where
\bea
\label{MB3}
U_0&=&\frac{1}{\xi-\xi_0}\Bigg(\sum_{l=1}^N A_l-diag \sum_{l=1}^N A_l\Bigg),\nn\\
\hat\rho &=& 4 N_0 \Bigg((\xi-\xi_0)\sigma_3-\sum_{l=1}^N \frac{A_l}{z_l}\Bigg).
\eea
Eqs.~(\ref{MB2}) admit a Hamiltonian structure with the Poisson brackets \cite{Kitaev}
\bea
\label{Poisson}
\big\{(A_m)_{ab},(A_n)_{cd}\big\}=\delta_{mn}\big((A_m)_{ad}\delta_{bc}-\delta_{ad}(A_m)_{bc}\big),\qquad
\eea
which corresponds to the ${sl}(2)$ algebra on a chain. Therefore, Eqs.~(\ref{MB2}) acquire the form
$\partial_{\xi}A_j=\big\{A_j, H_{\xi}\big\},\qquad \partial_{\eta}A_j=\big\{A_j, H_{\eta}\big\}$,
with Hamiltonian operators
\bea
\label{Ham}
H_{\xi}&=&\sum_{k=1}^N\left(z_k-\frac{\gamma_{\perp}}{2}\right)\tr(A_k \sigma_3)+\frac{S^-S^+ + S^+S^-}{\xi-\xi_0},\\
H_k&=&\sum_{k=1}^N\left(\frac{(\xi-\xi_0)}{z_k}-2 \gamma \right)\tr(A_k \sigma_3)+\sum_{j=1}^N\frac{\tr(A_kA_j)}{z_k(z_k-z_j)},\nn
\eea
and $S^{+,-}=(\sum_{k=1}^N A_k)_{12, 21}$.

Quantization of the MB system implies replacement of Poisson brackets by commutators, $\{, \}\rightarrow [,]$, and introduction of
a quantum wave function, $\Psi(\xi, \{z_k\})$. Then the $sl(2)$
algebra (\ref{Poisson}) acquires the following matrix realization
\bea
\label{algebra}
A_k = i \left(\begin{array}{cc}
S_k^3 & S_k^+\\
S_k^- & -S_k^3
            \end{array}
\right),
\eea
which, together with transformation $\xi \rightarrow i \xi$, leads to the set of Hamiltonian operators corresponding to (\ref{MBEqs}):
\bea
\label{Ham2}
\hat{h}_{\xi}&=& i^2 \sum_{k=1}^N\left(z_k-\frac{\gamma_{\perp}}{2}\right) \hat{S}^3_k+i^2\frac{\hat{S}^-\hat{S}^+ + \hat{S}^+\hat{S}^-}{\xi-\xi_0},\nn\\
\hat{h}_k&=&i^2\Bigl[(\xi-\xi_0)-2 \gamma z_k\Bigr]\hat{S}^3_k-i^2\sum_{j\neq k}^N\frac{\hat{\bf S}_k\hat{\bf S}_j}{z_k-z_j},\;\;
\eea
where $k=1\ldots N$, and $\hat{S}^{\pm}=\sum_k\hat{S}^{\pm}_k$. On a quantum level, the ``wave function,'' $\Psi(\xi, z_1\ldots z_{N})$, satisfying the set of ``multitime,''
$t\rightarrow (\xi, z_1\ldots z_N)$, Schr\"{o}dinger equations,
\bea
\label{multiS}
i \partial_{\xi}\Psi &=& \hat{h}_{\xi}\Psi, \nn\\
i \partial_{z_k}\Psi &=& \hat{h}_{k}\Psi,
\eea
unambiguously determine the solution of MB equations.
As we see, the second set of Schr\"{o}dinger equations coincides with generalized KZ equations (\ref{KZ2}) with the following
parameters:  $k+2=1$,  $i (\xi-\xi_0)= 1/\lambda$, $q(z)=-z/\lambda-\gamma z^2$, and $\cal{C}=\cal{C}_{\infty}$,
which encompasses all $z_j=(2 {\mathcal N}_0 \eta-k_j^2)^{1/2}$, where $k_j^2, \xi_0 \in \RR$ are free parameters.
The first equation is formally an ordinary Schr\"{o}dinger equation with ``time,'' $\xi=\Omega(t-x/c)$.
Therefore, this maps the problem onto a dynamical boundary WZNW model,  with the boundary action
$S_{MB}=\oint_{\mathcal{C}_{\infty}} dz q(z) J^3(z)$, correlation functions of which depend on an
additional parameter, $\xi$, playing the role of time.

By analogy with Eq.~(\ref{solution}), solution for $\Psi(\xi, z_1\ldots z_{N})$ is then found to be
\bea
\label{solution-3}
\Psi(\xi, z_1 \cdots z_{N})&=&\oint \prod_{k=1}^{M}du_k
\chi_{MB}(\{u_{\alpha}\}|\{z_i\})\nn\\
&\times&{\cal V}(\{u_{\alpha}\}|\{z_i\}),\;\;\qquad
\eea
with

\bea
\label{sol1}
\frac{\chi_{MB}(\{u_{\alpha}\}|\{z_i\})}{\chi_0(\{u_{\alpha}\}|\{z_i\})}= (\xi-\xi_0)^{i (\sum m_i- M)} e^{2 i (\xi-\xi_0)\sum z_i m_i}\nn\\
\times \exp\Bigl\{\frac{1}{\lambda}\Bigl[\sum_{i=1}^N m_i(\frac{\gamma_{\perp}}{2}-\lambda q(z_i))
+\sum_{\alpha} u_{\alpha}\Bigr]\Bigr\}.\qquad \qquad
\eea
Eqs.~(\ref{sol1}) and (\ref{solution-3}) determine the form of the wave function of quantum states in the quantized MB system. Importantly, this wave function is cardinally different from the correlation function of primary fields in the
{\em bulk} WZNW model. This is because the expression in the right-hand-side of Eq.~(\ref{sol1}) must be integrated in Eq.~(\ref{solution-3}) together with $\chi_0$, while only the later appears in the WZNW model.

In order to find the solution of the system of classical
MB equations (\ref{multiS}), one should average the angular momentum operators, $S^{\pm}$, $S^3$, with respect to the wave functions (\ref{sol1}).
Then the solution of classical MB equations
for physical quantities ${\mathcal E}$, ${\mathcal P}$, and ${\mathcal N}$ can be found from Eqs.~(\ref{MB22}), (\ref{MB2}), and  (\ref{MB3}) as follows:
\begin{eqnarray}
\label{solMBE}
\nonumber
&&{\mathcal E}^2=\left\langle\Psi^*  \Biggl|  \frac{\kappa^2}{(\xi-\xi_0)^2}\sum_{j=1}^N \hat{S}_j^- \sum_{j=1}^N \hat{S}_j^+ \Biggr| \Psi\right\rangle,\\
&&{\mathcal P}^2=64 {\mathcal N}^2_0 \kappa \left\langle\Psi^*  \Biggl| \sum_{j=1}^N \frac{\hat{S}^-_j}{z_j}\sum_{j=1}^N\frac{\hat{S}^+_j}{z_j}  \Biggr| \Psi\right\rangle,\\
&&{\mathcal N}=8 {\mathcal N}_0 \left\langle\Psi^* \Biggl| \Bigr[\xi-\xi_0-\kappa \sum_{j=1}^N\frac{{\hat{S}^3_j}}{z_j} \Bigr] \Biggr| \Psi\right\rangle.
\nonumber
\end{eqnarray}
These equations provide $N$-soliton solutions to the MB system. In general these solutions have compact integral representations which can be evaluated and compared with other numerical~\cite{num} and experimental~\cite{exp} data.
In  Appendix we evaluate this integral
for $N=2$ soliton case and express the solution for ${\mathcal E}$, ${\mathcal P}$, and ${\mathcal N}$  in terms of known Kummer confluent hypergeometric functions.

\section{Conclusion}
In conclusion, this work has introduced a method of reverse construction of boundary WZNW models
from the generalized Knizhnik-Zamolodchikov equations satisfied by the exact correlation functions
and demonstrated the application of this method on the explicit example of KZ equations with conserved
currents of the Richardson model. Thereby, we established a direct connection between the discrete pairing model of
superconductivity and the boundary WZNW model, which we identified and solved. We have established that the solutions
of modified KZ equations are defined by the off-shell states of the Richardson model. Our construction is close
in spirit but technically different from  the BCS/CFT correspondence discussed earlier by Sierra~\cite{Sierra1}.
Our main motivation has been to precisely identify the boundary operator in the WZNW model,
which is related to the Richardson-type models.
Our other motivation has been to outline a range of practical applications of the discovered correspondence,
which is argued to be very wide and includes a variety of dynamical systems that can be mapped on the BWZNW theories
and solved exactly in many cases. One such mapping and solution for the dynamical system describing radiation of a two level laser
with pumping and damping was presented. Dynamic properties of the laser were computed exactly exploiting the integrability of
this latter system.

{\em Acknowledgements --} This research was supported by the NSF CAREER award, DMR-0847224.

\section{Appendix}

Here we derive the analytical expression of the two-soliton solution of MB equations, when $N=2$
and $M=1$. For simplicity we consider the case with finite pumping, ${\cal N}_0$, and medium polarization damping, $\gamma_{\perp}$,
coefficients, but with $\gamma=0$.
In this case we have one integration  parameter $u$ and two parameters, $z_1, z_2$. Then from Eqs.~(\ref{sol-pm}) and (\ref{VS}) it follows that
\bea
\label{A1}
{\cal V}(u)&=&\sum_{i=1}^{2}\frac{\hat S_i^\dagger}{u-z_i}\mid\downarrow,\downarrow \rangle\nn\\
&=&\frac{1}{u-z_1}\mid\uparrow,\downarrow \rangle + \frac{1}{u-z_2}\mid\downarrow,\uparrow \rangle.
\eea
In order to construct the wave function  (\ref{solution-3}) and (\ref{sol1}),  we should integrate
${\cal V}(u)$  together with the term corresponding to $\chi_0$, $[(u-z_1)(u-z_2)]^{-1/2(k+2)}$, over $u$
along a contour surrounding
the brunch-cut at $(z_1,z_2)$ [see Eq.~(\ref{solution}) and the discussion].
To perform the resulting integration, we make use of the identity
\bea
\label{A2}
&&\oint_C du (u-z_1)^{-a} (u-z_2)^{-1+a}e^{-b u}\nn\\
&=&B(a,1-a)\text{\tiny{1}}\large{F}\text{\tiny{1}}\left[1-a; 1; b(z_2-z_1)\right]
\eea
where $\text{\tiny{1}}\large{F}\text{\tiny{1}}$ is the Kummer confluent hypergeometric function and
$B(a,1-a)$ is the Beta-function \cite{GR}. Then for  $a=1/2$ and $b=1/\lambda$ we have
\bea
\label{A21}
\text{\tiny{1}}\Large{F}\text{\tiny{1}}\left[\frac{1}{2}; 1; \frac{(z_1-z_2)}{\lambda}\right]=e^{\frac{z}{2\lambda}}I_0\left[\frac{(z_1-z_2)}{2\lambda}\right],
\eea
with $I_0$ being the modified Bessel function of zero order. Now, returning to the real time, $\xi \rightarrow -i \xi$,
by analytic continuation, and keeping causal
behavior of the wave function, $\Psi$, we will have
\begin{widetext}
\bea
\label{A3}
\Psi &=&\pi(\xi-\xi_0)^{-i}e^{-|\xi_-\xi_0|\frac{\gamma_{\perp}}{2}-|(\xi-\xi_0)(z_1-z_2)|}
\Biggl((z_1-z_2)^{-\frac{1}{4}}
e^{\frac{i(\xi-\xi_0)(z_2-z_1)}{2}}I_0\left[\frac{i(\xi-\xi_0)(z_2-z_1)}{2}\right]\mid\uparrow,\downarrow \rangle\nn\\
&+&(z_2-z_1)^{-\frac{1}{4}}
e^{\frac{i(\xi-\xi_0)(z_1-z_2)}{2}}I_0\left[\frac{i(\xi-\xi_0)(z_1-z_2)}{2}\right]\mid\downarrow,\uparrow \rangle\Biggr).
\eea

According to Eqs.~(\ref{solMBE}), to find ${\cal E}$ one should act by the operator $S^+_1+S^+_2$
on the expression (\ref{A3}) for $\Psi$
and calculate the norm.
By doing so and after some simple algebra we obtain
\bea
\label{A4}
{\cal E}&=&\frac{\pi}{|\xi-\xi_0|}e^{-|\xi_-\xi_0|\frac{\gamma_{\perp}}{2}-|(\xi-\xi_0)(z_1-z_2)|}\text{Abs}\Bigg[(z_1-z_2)^{-\frac{1}{4}}
e^{\frac{i(\xi-\xi_0)(z_2-z_1)}{2}}I_0\Big[\frac{i(\xi-\xi_0)(z_2-z_1)}{2}\Big]\nn\\
&+&(z_2-z_1)^{-\frac{1}{4}}
e^{\frac{i(\xi-\xi_0)(z_1-z_2)}{2}}I_0\Big[\frac{i(\xi-\xi_0)(z_1-z_2)}{2}\Big] \Bigg]
\eea
where the notion $\text{Abs}$ means absolute value. Similarly, the expressions for the polarization of the
medium  ${\cal P}$ and the population inversion $\cal N$ read:
\bea
\label{A5}
{\cal P} &= &8 \pi {\cal N}_0 e^{-|\xi_-\xi_0|\frac{\gamma_{\perp}}{2}-|(\xi-\xi_0)(z_1-z_2)|} \text{Abs}\Biggl[\frac{1}{z_2} (z_1-z_2)^{-\frac{1}{4}}
e^{\frac{i(\xi-\xi_0)(z_2-z_1)}{2}}I_0\left[\frac{i(\xi-\xi_0)(z_2-z_1)}{2}\right]\nn\\
&+&\frac{1}{z_1} (z_2-z_1)^{-\frac{1}{4}}
e^{\frac{i(\xi-\xi_0)(z_1-z_2)}{2}}I_0\left[\frac{i(\xi-\xi_0)(z_1-z_2)}{2}\right]\Biggr]
\eea
and
\bea
\label{A6}
{\cal N}&=&8 \pi {\cal N}_0 e^{-|\xi_-\xi_0|\frac{\gamma_{\perp}}{2}-|(\xi-\xi_0)(z_1-z_2)|}
Re\Biggl[\left(\xi-\xi_0+\frac{1}{2 z_2}-\frac{1}{2 z_1}\right) (z_1-z_2)^{-\frac{1}{4}}
e^{\frac{i(\xi-\xi_0)(z_2-z_1)}{2}}I_0\left[\frac{i(\xi-\xi_0)(z_2-z_1)}{2}\right]\nn\\
&+&\left(\xi-\xi_0+\frac{1}{2 z_1}-\frac{1}{2 z_2}\right) (z_2-z_1)^{-\frac{1}{4}}
e^{\frac{i(\xi-\xi_0)(z_1-z_2)}{2}}I_0\left[\frac{i(\xi-\xi_0)(z_1-z_2)}{2}\right] \Biggr].
\eea
\end{widetext}
We remind the reader that in all expressions above
$z_j=\sqrt{2 {\cal N}_0 \eta-k_j^2},\;\;( j=1,2)$, with constant parameters $k_1$ and  $k_2$.
Here, $\eta=\Omega x/c$, $\xi=\Omega(t-x/c)$ with $\Omega$ being a physical constant that characterizes
the material and the cavity medium. To the best of our knowledge, Eqs.~(\ref{A4}), (\ref{A5}) and (\ref{A6})
have never been derived in the literature before.



\begin{thebibliography}{99}

\bibitem{Afl} I. Affleck and F. D. M. Haldane, Phys. Rev. B {\bf 36}, 5291 (1987).

\bibitem{ludwig} S. Ryu, C. Mudry, A. Ludwig, and A. Furusaki, Nucl. Phys. B 839, 341-376 (2010).

\bibitem{PW} A. M. Polyakov and P. B. Wiegmann, Phys. Lett. B {\bf 131}, 121 (1983); {\bf 141}, 223 (1983).

\bibitem{KZ} D. Knizhnik and A. B. Zamolodchikov, Nucl. Phys. B{\bf 247}, 83 (1984).

\bibitem{Gaudin} M. Gaudin, J. Phys. {\bf 37}, 1087 (1976).

\bibitem{Sierra_RMP} J. Dukelsky, S. Pittel, and G. Sierra, Rev. Mod. Phys. {\bf 76}, 643 (2004)

\bibitem{R1} R. W. Richardson, Phys. Lett. {\bf 3}, 277 (1963);
R. W. Richardson, J. Math. Phys.  {\bf 18}, 1802 (1977).

\bibitem{Delft1} J. von Delft and D. C. Ralph, Phys. Rep. {\bf 345},
61 (2001) and references therein.

\bibitem{italians} M. C. Cambiaggio, A. M. F. Rivas, and M. Saraceno,
Nucl. Phys. A {\bf 624}, 157 (1997).

\bibitem{Amico} L. Amico, G. Falci, and R. Fazio, J. Phys. A {\bf 34}, 6425 (2001).

\bibitem{amico2} L. Amico and A. Osterloh, Phys. Rev. Lett. {\bf 88}, 127003 (2002).

\bibitem{Delft2}  J. von Delft and R. Poghossian, Phys. Rev. B {\bf 66}, 134502 (2002).

\bibitem{Sierra1} G. Sierra, Nucl. Phys. B {\bf 572}, 517 (2000);

\bibitem{Sierra2} G. Sierra,
Proc. of NATO Workshop on Stat. Field Theories,
Como 2001, hep-th/0111114.

\bibitem{teodorescu} R. Teodorescu, in {\em Leading-edge Superconductivity Research Developments.} Editor T. Watanabe.
Nova Science Publishers (2008).


\bibitem{Yuzbash} E. Yuzbashyan, A. Baytin, and B. Altshuler,
Phys. Rev. {\bf B 71}, 094505 (2005);
E. Yuzbashyan, Phys. Rev. {\bf B 78}, 184507 (2008).

\bibitem{garcia} A. M. Garcia-Garcia, J. D. Urbina, E. A. Yuzbashyan, K. Richter, and B. L. Altshuler,
Phys. Rev. Lett. {\bf 100}, 187001 (2008); arXiv:0911.1559.


\bibitem{caux} A. Faribault, P. Calabrese, and J.-S. Caux,
Phys. Rev. B {\bf 77}, 064503 (2008).

\bibitem{Ralph} D.C.Ralph, C.T. Black, and M. Tinkham,
Phys. Rev. Lett. {\bf 76}, 688 (1996); {\bf 78}, 4087 (1997).


\bibitem{CFT} A. A. Belavin, A. M. Polyakov, and A. B. Zamolodchikov, Nucl. Phys. B {\bf 241}, 333 (1984).





\bibitem{Babujian} H. M. Babujian, J. Phys. A {\bf 34}, 6425 (1994);
H. M. Babujian and R. Flume, Mod. Phys. Lett. A {\bf 9}, 2029 (1994).

\bibitem{gogolin} A. O. Gogolin, A. A. Nersesyan, and A. M. Tsvelik, {\em Bosonization in Strongly Correlated Systems.} Cambridge University Press (1999).

\bibitem{Aomoto} K.  Aomoto,  J. Math. Soc. Japan {\bf 39}, 191 (1987).

\bibitem{Gelfand} V. A. Vasilev, I. M. Gelfand and A. V. Zelevinski, Func. Anal. Priloz. {\bf 21}, 19 (1987).

\bibitem{Konik} P. Fendley and R. M. Konik,
Phys. Rev. B {\bf 62}, 9359 (2000).

\bibitem{lasers} T. Numai, {\em Fundamentals of Semiconductor Lasers.} Springer (2004).

\bibitem{BZM} S. P. Burtsev, V. E. Zakharov, and A. V. Mikhailov, Theor. Math. Phys. {\bf 70}, 227 (1987).

\bibitem{Kitaev} H. Babujian and A. Kitaev, J. Math. Phys. {\bf 39}, 2499 (1998).

\bibitem{Its} A. R. Its, Math. USSR Izv. {\bf 26}, 497 (1986).


\bibitem{num}
I. R. Al’miev, O. Larroche, D. Benredjem, J. Dubau, S. Kazamias, C. Möller, and A. Klisnick,
Phys. Rev. Lett. {\bf 99}, 123902 (2007); C. M. Kim, J. Lee, and K. A. Janulewicz, Phys. Rev. Lett. {\bf 104}, 053901 (2010).

\bibitem{exp} see {\em e.g.} Ph. Zeitoun, G. Faivre, S. Sebban, T. Mocek, A. Hallou, M. Fajardo, D. Aubert, Ph. Balcou, F. Burgy, D. Douillet, S. Kazamias, G. de Lachèze-Murel, T. Lefrou, S. le Pape, P. Mercère, H. Merdji, A. S. Morlens, J. P. Rousseau, and C. Valentin, Nature (London) 431, 426 (2004);
B. Cros, T. Mocek, I. Bettaibi, G. Vieux, M. Farinet, J. Dubau, S. Sebban, and G. Maynard, Phys. Rev. A 73, 033801 (2006);
J. Costello, Nat. Photon. 2, 67 (2008);
Y. Wang, E. Granados, F. Pedaci, D. Alessi, B. Luther, M. Berrill, and J. J. Rocca, Nat. Photon. 2, 94 (2008).

\bibitem{GR} I. S. Gradshtein and I. M. Ryzhik,   {\em Tables of integrals, series and products,}
Academic Press, INC (1996).



\end{thebibliography}
\end{document}